\documentclass[12pt]{article}
\usepackage{epsf,latexsym,rotate}
\usepackage{amsfonts,amssymb} %%amsmath
\epsfverbosetrue
\usepackage[ps,arc,frame]{xy}
\usepackage{graphicx}

\newcommand{\lapprox}{%
\mathrel{%
\setbox0=\hbox{$<$}
\raise0.6ex\copy0\kern-\wd0
\lower0.65ex\hbox{$\sim$}
}}
\newcommand{\gapprox}{%
\mathrel{%
\setbox0=\hbox{$>$}
\raise0.6ex\copy0\kern-\wd0
\lower0.65ex\hbox{$\sim$}
}}

\newcommand{\double}[1]{\mathbb{#1}}

\newcommand{\rr}{\double{R}}

\newcommand{\aaa}{\mathcal{A}}

\newcommand{\dd}{\mathcal{D}}

\newcommand{\hh}{\mathcal{H}}

\newcommand{\de}{\hbox{\rm{d}}}

\newcommand{\bb}{\begin{eqnarray}}
\newcommand{\ee}{\end{eqnarray}}
\newcommand{\eee}{\nonumber\end{eqnarray}}
\newcommand{\qq}{\quad}
 
\begin{document}

\font\twelve=cmbx10 at 13pt
\font\eightrm=cmr8

\thispagestyle{empty}

\begin{center}

CENTRE DE PHYSIQUE TH\'EORIQUE $^1$ \\ CNRS--Luminy, Case
907\\ 13288 Marseille Cedex 9\\ FRANCE\\

\vspace{2cm}

{\Large\textbf{Spectral action and big desert}} \\

\vspace{1.5cm}

{\large Marc Knecht $^2$, Thomas Sch\"ucker $^3$}

\vspace{2cm}

{\large\textbf{Abstract}}
\end{center}
The values of the Higgs mass are obtained for two possibilities of extending the standard model 
in a way compatible with the existence of a noncommutative structure at high energies. We assume 
the existence of a big desert between the low energy electroweak scale and the high energy scale
$\Lambda = 1.1\times 10^{17}$ GeV, where noncommutative features become relevant.  We conclude that it is extremely 
difficult to depart from the  Higgs mass value $ m_H= 175.1^{+5.8}_{-9.8}$ GeV 
obtained from noncommutative geometry for the standard model with three generations only.

\vspace{1.2cm}

\vskip 1,0 truecm

\noindent
PACS-92: 11.15 Gauge field theories\\
 MSC-91: 81T13 Yang-Mills and other gauge theories

\vskip 1truecm

\noindent CPT-P08-2006\\

\vspace{2cm}
\noindent $^1$ Unit\'e Mixte de Recherche  (UMR 6207)
 du CNRS  et des Universit\'es Aix--Marseille 1 et 2 et  Sud
Toulon--Var, Laboratoire affili\'e \`a la FRUMAM (FR 2291)\\
$^2$  knecht@cpt.univ-mrs.fr \\
$^3$ also at Universit\'e Aix--Marseille 1, 
 schucker@cpt.univ-mrs.fr 

\section{Introduction}

Noncommutative geometry \cite{book} in its mild version of almost commutative geometry allows to derive certain Yang-Mills-Higgs models from general relativity through the spectral action principle \cite{real,grav,cc}. The standard model of electromagnetic, weak and strong forces is among them, and this derivation unifies non-Abelian gauge- and Higgs-bosons. As a consequence, their coupling constants are related. In the case of three
generations of leptons and quarks, this relation reads
\bb g_2^2=g_3^2= \lambda /8 .\label{2constr}
\ee
Relations of this type face two problems:
\begin{itemize}\item
The first equation is in flagrant contradiction with the experimental values of the weak and strong gauge couplings at accessible energies.
\item
Due to short distance divergencies in quantum corrections, the coupling constants are scale dependent,
and these two relations are not preserved under the renormalisation group flow.
Thus, they can at best only hold at a given energy scale.
\end{itemize}

 The first equation, $g_2=g_3$, is familiar from grand unified theories \cite{gut} where it comes from an embedding of the standard-model group $ SU(3)\times SU(2)\times U(1)$ into a simple group like $SU(5)$. In grand unified theories, the first problem is solved by the hypothesis of the {\it big desert}, which extends from present high energies all the way up to the astronomical energy of $\Lambda\sim 10^{17} $ GeV. One assumes
\begin{itemize}\item
  that -- with the exception of the Higgs scalar -- no new particles  exist with masses in the range of energies covered by the big desert, 
 \item 
  that perturbative quantum field theory remains valid throughout the desert.
\end{itemize}
Then the renormalisation group flow remains unchanged all the way up from the low energy electroweak scale to the scale $\Lambda $ where the weak and strong gauge couplings become indeed equal. At energies above $\Lambda $, one still faces the second problem, which is solved by a second hypothesis: one introduces new particles, {\it lepto-quarks}. They are the gauge bosons of $ SU(5)/[ SU(3)\times SU(2)\times U(1)]$ and are given  a mass equal to $\Lambda $ via spontaneous symmetry breakdown induced by a second Higgs multiplet in the adjoint representation. Then, by the Appelquist-Carazzone decoupling theorem \cite{ac}, the lepto-quarks do not upset the renormalisation flow in the desert, but do stabilize the equality $g_2=g_3$ at energies above $\Lambda $. However, in spite of their huge mass, the lepto-quarks render the proton unstable with a life-time that contradicts today's experimental numbers.

Let us now keep only the first hypothesis, the existence of a big desert, and let us assume that at the higher end of the desert it is not the particle spectrum that changes, but the geometry of spacetime itself. We shall assume that the (commutative) Riemannian geometry of spacetime is only a low energy approximation of a -- not yet known -- noncommutative geometry. Being noncommutative, this geometry has radically different short distance properties and is expected to produce quite a different renormalisation flow. It even might have no short distance divergencies at all, and therefore constant coupling constants, i.e. vanishing $\beta $-functions. At energies below $\Lambda $, this noncommutativity manifests itself only in its mild, almost commutative version through the gauge- and Higgs-fields of the standard model, which are magnetic-like fields accompanying the gravitational field. The first example of a truly noncommutative geometry is the noncommutative torus \cite{torus} or its non-compact version, the Moyal plane \cite{moyal}. The spectral action may be defined on theses spaces \cite{gayral}, leading to quantum field theories with non-local interactions \cite{star}. The best studied example is Grosse and Wulkenhaar's  scalar $\varphi ^4$ theory on the the Moyal $\rr^4$. This theory has two remarkable properties: it is renormalizable to all orders of perturbation theory \cite{gw1} and in the self-dual version, its one-loop $\beta $-functions indeed vanish \cite{gw2}.

These considerations motivate the combination of the two  constraints (\ref{2constr}) from almost commutative geometry \cite{cc}, that are assumed to hold at the scale $\Lambda$ (and perhaps above $\Lambda$),  with the renormalisation group flow of the standard model \cite{mv}, that remains relevant below $\Lambda$. From reference  \cite{cmpp} we learn -- and this is not at all obvious -- that this combination is possible without leaving the perturbative domain (`triviality') and without running into negative Higgs self-couplings (`stability'). To one loop, if all fermion masses except the top mass as well as threshold effects are neglected, the resulting value of the Higgs mass lies in the range \cite{mrs}
\bb m_H= 175.1\matrix{+5.8\cr -7.2} \ {\rm GeV\qq with \qq}
m_t= 178.0\ \pm\ 6.0\ {\rm GeV}.\ee

The aim of this paper is to investigate how the Higgs mass changes if we try to allow for new particles in the desert. Of course, the blooming of the desert is an old question and has many answers, depending on which new particles one chooses. In almost commutative geometry however, this choice is extremely constrained. Section 2 presents these constraints and two possible choices are displayed, for which the corresponding Higgs masses are computed in section 3.

\section{Did you say bloom?}

A noncommutative geometry is defined by a spectral triple $(\aaa,\hh,\dd)$ where $\aaa$ is a pre $C^*$-algebra represented on a Hilbert space $\hh$ on which also the Dirac operator $\dd$ acts. The spectral action is an invariant action defined on the spectral triple. The invariance is under all algebra automorphisms and is naturally ensured by making use only of the spectrum of the Dirac operator. In the commutative case (Riemannian geometry), the Hilbert space consists of square integrable spinors, the Dirac operator encodes the metric, the algebra automorphisms are just general coordinate transformations and the spectral action reproduces the Einstein-Hilbert action.

In the almost commutative case, the algebra automorphisms determine in addition the (non-Abelian part of) the gauge group, the Hilbert space gives in addition the fermion gauge-multiplets whose masses are found in the Dirac operator. The spectral action then yields the Yang-Mills action, the Higgs representation and its symmetry breaking Higgs potential.

Krajewski, Paschke and Sitarz \cite{kps} have classified all almost commutative geometries. This classification is conveniently organized by Krajewski diagrams, which are the analogues of Dynkin and weight diagrams in group theory. Through the abundant use of Krajewski diagrams, the classification can then be further reduced with the help of a list of physics motivated requirements \cite{refine}. Two of them are borrowed from grand unified theories: the spectral triple is taken irreducible and its Hilbert space is to define a {\it complex } representation under the little group. Other requirements concern the physical properties of the ensuing Yang-Mills-Higgs model, like the vanishing of anomalies, or the possibility to allow for non-degenerate masses in irreducible fermion multiplets. The surviving almost commutative geometries are scarce. There is no solution if the algebra is simple, or if it has two simple summands. With three simple summands, there only remains the standard model with an arbitrary number of colours $p$ and with one generation\footnote{The standard model with an arbitrary number
of generations corresponds to a {\it reducible} spectral triple.}
of leptons and quarks. In addition, the neutrino must be massless. A few submodels like $SU(p)\times SO(2)\times U(1)$ or $SO(p)\times SU(2)\times U(1)$ are also possible. With four summands, only one more model survives, the $SU(p)\times U(1)$ electro-strong model with one quark and one lepton. This model is vector-like, has no Higgs and the fermion masses are gauge-invariant.

This reduced classification explains the empirical fact that it is extremely difficult to find viable extensions of the standard model in almost commutative geometry. One obvious possibility is a fourth generation of leptons and quarks with a massive neutrino.

Another possibility coming from an algebra with six summands has been found recently \cite{chris}. It is identical to the standard model with two additional leptons $A^{--}$ and $C^{++}$ whose electric charge is two in units of the electron charge. These new leptons  couple neither to the charged gauge bosons, nor to the Higgs scalar. Their hypercharges are vector-like, so that they do not contribute to the electroweak gauge anomalies. Their masses are gauge-invariant and they constitute viable candidates for cold dark matter \cite{klop}.

Before proceeding, let us quote here the present experimental constraints on the masses of a fourth generation \cite{data}, whose quarks are denoted by $(T,B)$ and the leptons by $(N,E)$:
\bb m_B>199\ {\rm GeV},\ m_N>45.6\ {\rm GeV},\ m_E>100.8\ {\rm GeV},\  
\Delta m_Q^2+
{\textstyle\frac{1}{3}} \Delta m_L^2<(85\ {\rm GeV})^2, \label{4ex}\ee
with
\bb \Delta m_Q^2:=m_T^2+m_B^2-\frac{4m_T^2m_B^2}{m_T^2-m_B^2}\,\ln\frac{m_T}{m_B}\,,\qq
\Delta m_L^2:=m_N^2+m_E^2-\,\frac{4m_N^2m_E^2}{m_N^2-m_E^2}\,\ln\frac{m_N}{m_E}\,.\ee
We shall see in the next section that these experimental constraints are incompatible with a perturbative running of the Yukawa couplings in the fourth generation up to $\Lambda $.

\section{The renormalisation group flow}

We normalize the Higgs self-coupling as $(\lambda /24)\, \varphi ^4$. Then 
\bb m_H^2=\,\frac{4}{3}\,\frac{\lambda }{g_2^2}\,m_W^2\ee
at tree level.
The hypercharge coupling $g_1$ is the one of the Glashow-Salam-Weinberg model (without the  factor $\sqrt{5/3}$ from grand unified theories).
The Yukawa couplings are normalized such that
\bb m_f=\sqrt{2}\,\frac{g_f}{g_2}\,m_W.\ee
We further use:
\bb t:=\ln (\mu/m_Z),\qq \de g/\de t=:\beta _g,\qq \kappa :=(4\pi )^{-2},\ee
where $\mu$ denotes the renormalization group scale.

\subsection{Four generations}

Neglecting all fermion masses below the top mass, as well as all  fermion mixing effects, we have the following one-loop $\beta $-functions for the standard model with $N=4$ generations \cite{mv}:
\bb \beta _{g_i}&=&\kappa b_ig_i^3,\qq b_i=
{\textstyle 
\left( \frac{20}{9} N+\frac{1}{6},-\frac{22}{3}+\frac{4}{3} N+\frac{1}{6},
-11+\frac{4}{3} N\right) }
\\ \cr 
\beta _t&=&\kappa
 \left[ -\sum_i c_i^ug_i^2 +Y_2 +\,\frac{3}{2}\,g_t^2
 \,\right] g_t,\\
 \beta _T&=&\kappa
 \left[ -\sum_i c_i^ug_i^2 +Y_2 +\,\frac{3}{2}\,g_T^2
 -\,\frac{3}{2}\,g_B^2\right] g_T,\\
 \beta _B&=&\kappa
 \left[ -\sum_i c_i^dg_i^2 +Y_2 +\,\frac{3}{2}\,g_B^2
 -\,\frac{3}{2}\,g_T^2\right] g_B,\\
 \beta _N&=&\kappa
 \left[ -\sum_i c_i^\nu g_i^2 +Y_2 +\,\frac{3}{2}\,g_N^2
 -\,\frac{3}{2}\,g_E^2\right] g_N,\\
 \beta _E&=&\kappa
 \left[ -\sum_i c_i^eg_i^2 +Y_2 +\,\frac{3}{2}\,g_E^2
 -\,\frac{3}{2}\,g_N^2\right] g_E,\\
 \beta _\lambda &=&\kappa 
 \left[ \,\frac{9}{4}\,\left( g_1^4+2g_1^2g_2^2+3g_2^4\right) 
 -\left( 3g_1^2+9g_2^2\right) \lambda 
 +4Y_2\lambda -12H+4\lambda ^2\right] ,\ee
 with
 \bb c_i^u=\left( {\textstyle\frac{17}{12}},{\textstyle\frac{9}{4}} , 8\right) ,
 &
 c_i^\nu =\left( {\textstyle\frac{3}{4}},{\textstyle\frac{9}{4}} , 0\right) ,
 &
 Y_2=3g_t^2+3g_T^2+3g_B^2+g_N^2+g_E^2,
 \\
 c_i^d=\left( {\textstyle\frac{5}{12}},{\textstyle\frac{9}{4}} , 8\right) ,
 &
c_i^e=\left( {\textstyle\frac{15}{4}},{\textstyle\frac{9}{4}} , 0\right) ,
&
 H=3g_t^4+3g_T^4+3g_B^4+g_N^4+g_E^4.
\ee
We shall integrate the renormalisation  group equations  $\de g/\de t=\beta _g$ neglecting threshold effects. The three gauge couplings decouple from the other equations and give:
\bb g_i(t)=g_{i0}/\sqrt{1-2\kappa b_ig_{i0}^2t}.\ee
The initial conditions are taken from experiment \cite{data}:
\bb g_{10}= 0.3575,\qq
g_{20}=0.6514,\qq
g_{30}=1.221.\ee
The unification scale $\Lambda $ is the solution of $g_2(\ln (\Lambda /m_Z))=g_3(\ln (\Lambda /m_Z))$,
\bb \Lambda = m_Z\exp\frac{g_{20}^{-2}-g_{30}^{-2}}{2\kappa (b_2-b_3)}\,=\,1.1\times 10^{17}\  {\rm GeV},\ee
and is independent of the number of generations.
The equations for the Yukawa couplings decouple from  that of the Higgs self-coupling, and we integrate them numerically with initial conditions taken again at $t=0$, i.e. $\mu=m_Z$:
\bb m_t= 178.0\ {\rm GeV},\qq m_T=m_B=m_N=m_E=:m_4.\ee
This integration up to $\Lambda $ makes sense only if $m_4 <  105$ GeV, the Yukawa coupling $g_N$ being the first to leave the perturbative domain, 
\bb g_N(\ln\Lambda /m_Z)= 2.1,\qq m_4=105\ {\rm GeV}.\ee
Finally we integrate numerically the Higgs self-coupling with the final condition at $\mu=\Lambda $ taken from almost commutative geometry. For $N$ generations, this condition reads \cite{cc}
\bb g_2^2=\,\frac{3}{N}\,\frac{Y_2^2}{H}\,\frac{\lambda}{24}\,,\ee
and generalizes the second constraint of (\ref{2constr}).
From the Higgs self-coupling $\lambda (0)$ at the electroweak scale $\mu=m_Z$, we then infer the Higgs mass. Some numerical results are shown in Table 1. The largest uncertainty in the Higgs mass, $\pm 6.5$ GeV, comes from the experimental error in the top mass. The neglected threshold effects should not induce an uncertainty larger than $ 2.6$ GeV. Indeed if we evaluate the ratio $\lambda /g_2^2$ for three generations at 175.1 GeV rather than at $m_Z$, we find a Higgs mass of 172.5 GeV instead of 175.1 GeV.

\begin{table}[h]
\begin{center}
\begin{tabular}{|c|c|}
\hline\\[-3mm]
$m_4\ [{\rm GeV}]$&$m_H\ [{\rm GeV}]$\\[1ex]
\hline
\, 50&167.5\\
\ \,65&166.3\\
\ \,80&168.2\\
100&177.4\\
105&180.7\\
\hline
\end{tabular}
\end{center}
\caption{The Higgs mass as a function of a common mass in the 4th generation}
\end{table}

\begin{table}[h]
\begin{center}
\begin{tabular}{|c|c|c|c|c|c|}
\hline\\[-4mm]
$m_T\ [{\rm GeV}]$&
$m_B\ [{\rm GeV}]$&
$m_N\ [{\rm GeV}]$&
$m_E\ [{\rm GeV}]$&
$m_H\ [{\rm GeV}]$&
$g_Y$\\[1ex]
\hline
\, 50&\, 50&\, 50&100&171.6&$g_E$=0.75\\
\, 50&100&\, 50&\, 50&167.6&$g_N$=0.38\\
100&\, 50&\, 50&\, 50&167.6&$g_N$=0.38\\
\, 50&100&\, 50&100&172.3&$g_E$=0.89\\
\, 50&100&\, 50&100&172.3&$g_E$=0.89\\
100&100&\, 50&\, 50&168.9&$g_N$=0.43\\
100&100&\, 50&100&174.1&$g_E$=1.11\\
100&125&\, 50&100&179.5&$g_E$=1.50\\
125&100&\, 50&100&179.4&$g_E$=1.48\\
125&125&\, 50&100&184.4&$g_E$=2.82\\
125&150&\, 50&\, 50&190.5&$g_B$=1.07\\
150&125&\, 50&\, 50&190.4&$g_T$=0.99\\
\hline
\end{tabular}
\end{center}
\caption{The Higgs mass as a function of masses in the 4th generation, $g_Y$ is the largest Yukawa coupling
at the scale $\mu = \Lambda$.}
\end{table}

Table 2 collects a few results with mass splittings in the fourth generation. To show how close they are to the limit of the perturbative domain, the largest Yukawa coupling is also indicated. We have found no mass pattern in the fourth generation compatible with the four experimental constraints (\ref{4ex}) and compatible with perturbation theory up to $\Lambda $.

\subsection{Three generations plus $AC$-leptons}

Instead of a fourth generation, let us add Stephan's doubly charged $AC$-leptons \cite{chris} to the $N=3$ generations of the standard model. Since the $AC$-leptons carry only hypercharge, their presence only modifies $b_1$, which now becomes
\bb b_1=\frac{20}{9}\cdot 3+\frac{1}{6}+\frac{32}{3}.\ee
For simplicity, let us endow both $AC$-leptons with the same mass $m_{AC}$. There is little
experimental information about mass limits for doubly-charged heavy leptons from direct searches  
\cite{data}.
Such charged leptons would also contribute to low energy precision observables, like the anomalous
magnetic moment of the muon $a_\mu$. The lowest order contribution to $a_\mu$, which reads
\bb
\Delta a_\mu(A^{--}) = \Delta a_\mu(C^{++}) = \left( \frac{\alpha}{\pi}\right)^2\,B_2(m_{AC}),
\ee
%
%
%%%%%%%%%%%%%%%%%%%%%%%%%%%%%%%% Figure 12 %%%%%%%%%%%%%%%%%%%%%%%%%%%%%%%%%%%
\begin{figure}[h]
\vspace*{0.05in}
%\rule{5cm}{0.2mm}\hfill\rule{5cm}{0.2mm}
%%%%%\centerline{\epsfbox{xs9097.ps}}
\centering                                             % ajoute 
       \includegraphics[bb=150 450 380 590]{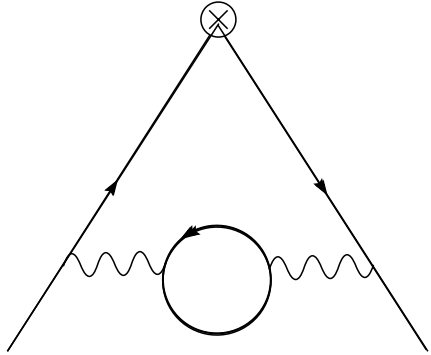}
%\rule{5cm}{0.2mm}\hfill\rule{5cm}{0.2mm}
\caption{The insertion of the vacuum polarization loop of $AC$-leptons
into the one-loop QED correction to the anomalous moment of the muon }
\end{figure}
%%%%%%%%%%%%%%%%%%%%%%%%%%%%%%%%%%%%%%%%%%%%%%%%%%%%%%%%%%%%%%%%%%%%%%%%%%%%%%

\noindent
occurs through
the insertion of a vacuum polarization loop of $AC$-leptons into the one-loop
photonic correction, see Figure 1. In this expression, we have \cite{a_mu}
\bb
B_2(m_{AC}) = (q_{AC})^2\,\frac{1}{3}\,\int_{4m_{AC}^2}^\infty
ds\,\sqrt{1-\frac{4m_{AC}^2}{s}}\,\frac{s + 2 m_{AC}^2}{s^2}\,
\int_0^1 dx\,\frac{x^2(1-x)}{x^2 + (1-x)s/{m_\mu^2}}.
\ee
For our purposes, we only need the leading behaviour of $B_2(m_{AC})$ for $m_{AC}\gg m_\mu$ 
\cite{LdR68},
\bb
B_2(m_{AC}) \sim (q_{AC})^2\,\frac{1}{45}\,\left( \frac{m_\mu}{m_{AC}}\right)^2 ,
\ee
where $q_{AC} = 2$ is the charge of the $AC$-leptons. To keep this contribution
below the uncertainty $\left( \Delta a_\mu\right)^{\mbox{\scriptsize{exp}}} = 6.3 \times 10^{-10}$
achieved by the high precision measurement of $a_\mu$ at BNL \cite{E821final}, 
requires $m_{AC}\gapprox 10$ GeV. Finally, constraints from cosmology
give an upper bound  $ m_{AC}\ \lapprox\ 10^7\ {\rm GeV}$ \cite{klop}.  

If the mass of the $AC$-leptons is too small, the gauge coupling $g_1(\mu)$, for instance,
does not stay in the perturbative regime throughout the desert. We have therefore restricted
the mass of the $AC$-leptons to the range $ 100\ {\rm GeV} \lapprox\ m_{AC}\ \lapprox\ 10^7\ {\rm GeV}$.
As shown in Table 3, it then follows that these leptons modify the Higgs mass only slightly. 
At the same time, they increase the value of $g_1$ at $\mu=\Lambda $. These values are to be compared to $g_1=0.46$ at $\mu = \Lambda $ for three generations and no new fermions. 

\begin{table}[h]
\begin{center}
\begin{tabular}{|c|c|c|}
\hline\\[-3mm]
$m_{AC}\ [{\rm GeV}]$&$m_H\ [{\rm GeV}]$&
$g_1$ at $\Lambda$\\[1ex]
\hline
$\Lambda$& 175.1&0.46\\
$10^7$& 175.3&0.77\\
$10^5$& 175.4&0.97\\
$10^3$& 175.5&1.49\\
$200$& 175.3&2.08\\
$130$& 175.0&2.40\\
$100$& 174.7&2.69\\
\hline
\end{tabular}
\end{center}
\caption{The Higgs mass in presence of $AC$-leptons}
\end{table}

\section{Conclusion}

Some very special Yang-Mills-Higgs models can be derived from general relativity by extending Riemannian geometry to almost commutative geometry. After imposing a few physically motivated conditions on these special models, we remain with the standard model including an arbitrary number of colours and generations and constraints on gauge- and Higgs-couplings.  These constraints are assumed to be valid at a very
high energy scale $\Lambda$ where the noncommutative aspects become visible. If we choose three colours and three generations, and if we also admit the hypothesis of the big desert, these constraints point towards a Higgs mass of $ m_H= 175.1^{+5.8}_{-9.8}$ GeV. In this paper we tried to weaken the hypothesis of the big 
desert in two different ways compatible both with almost commutative geometry and with {\it perturbative} 
one-loop corrections, and we have computed the corresponding consequences on the Higgs mass.

First we have introduced a fourth generation. The compatibility with perturbation theory implies that the masses of the new fermions must be small, and so are their effects on the Higgs mass. However the smallness of
these fermion masses is then in conflict with the present day constraints from high energy physics.

Second we considered Stephan's doubly charged $AC$-leptons. We have examined various constraints that can be put on their masses. The requirement that the evolution of the couplings remains perturbative
throughout the desert combined with constraints from cosmology requires their masses to
be between $10^2$ GeV and $10^{7}$ GeV. Their effect on the Higgs mass is then tiny.   

Thus, the value of the Higgs mass turns out to be a very stable quantity for the class of models
constrained by noncommutative geometry that we have considered here. 

\indent

\noindent
{\bf Acknowledgements}\\
The work of one of us (M. K.) is supported in part by the EC contract 
No. HPRN-CT-2002-00311 (EURIDICE).

\end{document}